\documentclass[12pt,preprint]{aastex}

\usepackage{epsfig}
\begin{document}

\title{Observational Signatures of Binary Supermassive Black Holes}

\author{Constanze Roedig\altaffilmark{1}, Julian H. Krolik\altaffilmark{1}}

\altaffiltext{1}{Department of Physics and Astronomy, Johns Hopkins University, Baltimore, MD 21218, USA}

\and

\author{M. Coleman Miller\altaffilmark{2}}

\altaffiltext{2}{Department of Astronomy and Joint Space-Science Institute, University of Maryland, College Park, MD 20742}

\begin{abstract}

Observations indicate that most massive galaxies contain a supermassive
black hole, and theoretical studies suggest that when such galaxies have a major
merger, the central black holes will form a binary and eventually coalesce.
Here we discuss two spectral signatures of such binaries that may help
distinguish them from ordinary AGN.   These signatures are expected when
the mass ratio between the holes is not extreme and the system is fed
by a circumbinary disk.   One such signature is a notch in the thermal continuum
that has been predicted by other authors; we point out that it should be
accompanied by a spectral revival at shorter wavelengths and also discuss
its dependence on binary properties such as mass, mass ratio, and separation.
In particular, we note that the wavelength $\lambda_n$ at which the notch
occurs depends on these three parameters in such a way as to make the number of systems
displaying these notches  $\propto \lambda_n^{16/3}$; longer wavelength searches
are therefore strongly favored.
A second signature, first discussed here, is hard X-ray emission with a Wien-like spectrum at
a characteristic temperature $\sim 100$~keV produced by Compton cooling of the
shock generated when streams from the circumbinary disk hit the accretion disks around the 
individual black holes.  We investigate the observability of both signatures.   The hard
X-ray signal may be particularly valuable as it can provide an indicator of black hole
merger a few decades in advance of the event.

\end{abstract}

\section{Introduction}

Binary supermassive black holes are objects of great intrinsic interest.  
That two massive black holes should coexist in a single
galaxy appears to be a natural corollary of the standard theory of galaxy
formation, in which galaxies are built up by mergers from smaller
galaxies (see \citealt{2012RAA....12..917S} for a recent review), and the
observation that nearly every galaxy whose luminosity is
more than a fraction of the characteristic galaxy luminosity $L_*$ contains a
supermassive black hole in its
nucleus \citep{KormendyHo2013}. Their progress toward ultimate merger,
however, depends upon the gas content and stellar orbital distribution of
the galaxy in which they live
\citep{BBR80,GouldRix2000,2003ApJ...596..860M,2004ApJ...607..765E,2005ApJ...630..152E,2006ApJ...642L..21B,Cuadra2009,2009JPhCS.154a2049P,2009ApJ...695..455B,2011ApJ...732...89K,Vasiliev13,2013ApJ...773..100K}.  
In addition, binary supermassive black holes should be prodigious sources
of gravitational wave emission during their actual merger, and detectable
by pulsar timing arrays even well before merger (see \citealt{2013MNRAS.433L...1S} for a recent calculation).

Despite considerable effort, observational examples are few and limited
in their character.  There are now a number of ``duals", which are two
supermassive black holes that are in the same galaxy but are not
gravitationally bound to each other in a true binary
\citep{Komossa03,Comerford13,Liu13,Woo13}.   These can be identified
through a number of techniques, including X-ray or optical imaging
revealing two AGN-like point-sources in a single galaxy and doubled line
profiles.   These methods require that both black holes receive a large
enough accretion flow to ``light up" as AGN.  Analogous spectral methods
have been employed to search for genuine binaries, but they are difficult
to use.   For example, if both members of the black hole binary are
AGN, and their separation is large enough that their mutual orbital
speeds are small compared to the width of broad emission lines, the
offset between their respective broad line profiles will be small
compared to their intrinsic widths and hard to discern; conversely, if
the separation is small enough to produce a larger orbital velocity, the
broad emission line regions of the two will overlap, and the lines will
respond to the combined gravitational potential and ionizing radiation of
both black holes \citep{ShenLoeb10,2012ApJS..201...23E}.    Alternatively, shifts in broad emission
line peaks or centroids may perhaps indicate binaries \citep{Loeb13,Decarli13}.

\cite{ArtymLubow94} showed that circumbinary disks around binary systems
with mass ratios $q \gtrsim 4 (h/r) \alpha^{1/2}$ have low density cavities around
the binary because stable circular orbits do not exist within a radius $\sim 2a$ of their
centers of mass (here $q = M_2/M_1$ for $M_1$ the larger mass and $M_2$ the
smaller; $h/r$ is the disk aspect ratio; $\alpha$ is the ratio of stress to pressure within the disk;
and $a$ is the binary's semimajor axis).    In the context of supermassive black
hole binaries, the lower bound on $q$ for creating a cavity can be quite small.
Although initial studies of circumbinary disk
dynamics suggested that little of the mass accreting through the outer
regions of the disk would pass through the inner edge at $r \simeq 2a$
and be captured by the binary \citep{Pringle91}, recent work has indicated that the ``leakage fraction"
is at least $\sim 10\%$ \citep{MM08,Shi2012,Roedig12,Farris13} and
possibly unity (Shi et al. 2014 in preparation).  All studies so far that have
examined the question find that the majority of the accretion goes to the secondary rather
than the primary when the mass ratio is not too small \citep{Roedig12,Farris13}.
This implies that accretion will increase the mass ratio. It also implies
that if the accretion rate fed to the circumbinary disk is enough to support a
luminous AGN, as is likely in a galaxy that is the product of a recent merger
\citep{Hopkins12,Mayer13}, at least one and possibly both of the black holes will acquire
its own accretion disk and appear as an AGN.

There has been much interest in finding distinctive electromagnetic signatures
of these systems, partially to supplement future gravitational wave observations,
but also so that they may be found and studied even before instruments capable
of detecting their gravitational waves are built.   Several papers have noted that the
absence of optically thick gas in the low-density region between the circumbinary disk and
the accretion disks around each black hole (from here on, called ``minidisks")
will produce a dip in the thermal continuum spectrum
\citep{Roedig12,Tanaka12,GM12,Kocsis12,Tanaka13}.    However, all but  one
\citep{Roedig12} suppose that minidisks either do not exist or receive matter at
a rate substantially below the accretion rate in the circumbinary disk, thus
underestimating the thermal continuum at frequencies above the dip, nor
has there been any study of the summed spectrum's dependence on parameters.
More importantly, all previous work (except \cite{Tanaka13b}, discussed later)
neglects the spectral contribution from
the energy released when the fluid streams that carry matter on ballistic
orbits from the inner edge of the circumbinary disk to the outer edges of
the individual disks shock and release their kinetic energy as their
contents join those disks.

Here we present a more complete account of these radiation processes and explore
how they depend on the binary parameters $M=M_1 + M_2$, $q$, and $a$.  We also
use estimates of binary lifetimes at their different evolutionary stages to predict which
signals should be most common and easiest to observe.  In Section~2 we discuss thermal radiation
from the circumbinary disk and the two minidisks.  In Section~3 we discuss
the previously neglected radiation from the hot spots produced by impact
of the ballistic streams onto the minidisks.  In Section~4 we explore the
observability of our signatures as a function of masses, mass ratios, and
separations, and we present our conclusions in Section~5.
 
\section{Thermal Disk Radiation}

There is some disagreement in the literature about whether minidisks will
form even if most or all of the accretion through the circumbinary material is
delivered to one or the other of the holes.
For example, \cite{Tanaka13} and \cite{Tanaka13b} argue that the mini-disks will be
transient because the angular momentum of the streams relative to the individual black
holes is so small that circularization will be achieved at a radius very small compared to the
binary separation, and internal stresses will be very strong because the gas will shock to
a temperature near the virial temperature.   They then predict that radiation from the
mini-disks would take the form of brief thermalized flares modulated on the binary orbital period.
However, as discussed in \citet{MK13}, if the average specific angular momentum of
the matter in the streams is greater than the specific angular momentum at
the innermost stable circular orbits (ISCOs) of the holes, the matter's angular momentum
must be reduced by transfer outward.    When the disk is small, the net effect of this transfer
is to push other mass outward, enlarging the disk until it reaches the tidal truncation radius.
Only when the disk extends that far can the outward angular momentum flux be efficiently
transferred to the orbit.
The only loophole to this argument opens when the
specific angular momentum of the streams is so small that it does not even support
a circular orbit at the ISCO, and the streams flow more or less directly into the black holes
\citep{Gold13}.   There are then no minidisks at all.  Such a situation is associated with
very small binary separations.   We conclude that, except for very closely separated
binaries, the minidisks will extend to their tidal truncation radii.   

The thermal radiation from an accreting supermassive black hole binary is then the
sum of the radiation from the circumbinary disk and
the radiation from the two minidisks.  The gas between the minidisks and the
circumbinary disk has low surface density and hence low emissivity. If the
binary mass ratio $q = M_2/M_1$ is not too much less than unity, the
inner edge of the circumbinary disk is generically located at $r \simeq 2a$ \citep{Shi2012,Farris13},
and the tidal truncation radii of both disks is less than $\sim a/2$
\citep{Paczynski77}.  Thus the radiation that in an ordinary disk would have
been radiated between the radii $\sim a/2$ and $2a$ is missing.  Because
most of the light at frequency $\nu$ radiated by an ordinary disk is
produced near the radius $r(\nu)$ where $kT \sim h\nu$, a dark ``notch" in
the spectrum is carved across a factor of at least several in frequency.  

Far from edges, the effective temperature of an accretion disk in inflow equilibrium is
\begin{equation}\label{eqn:teff}
kT_{\rm eff}=\left[{3\over{8\pi\sigma}}{GM{\dot M}\over r^3}\right]^{1/4}.
\end{equation}
Here $\sigma$ is the Stefan-Boltzmann constant, $M$ is the total mass of
the binary, and ${\dot M}$ is the total accretion rate.    This relationship
must be adjusted at disk edges where the nature of the accretion flow changes.
For example, at the inner edge of a circumbinary disk, although the magnetic
stress continues smoothly, Reynolds stress can increase significantly due to
the impact of streams that leave the edge, pass close enough to the binary
to be torqued to higher angular momentum, and are flung outward \citep{Shi2012}.
Such additional stresses can raise the effective temperature above that of an untruncated disk
\citep{Cuadra2009, Roedig12, Gold13}.
Alternatively, near the ISCO, the flattening of the circular orbit angular
momentum profile permits inflow with less internal stress, although the continued
operation of MHD stresses prevents a complete cessation \citep{KHH05,Noble10,Penna13}.
Despite these possible complications, we estimate the characteristic temperature of the
notch in terms of the temperature (as defined by Equation~(\ref{eqn:teff})) that would be achieved
in an accretion flow onto a single black hole with mass $M$ and accretion rate $\dot M$ at $r\sim a$.
We expect this temperature ($T_0$) to lie between the hottest temperature in the circumbinary disk and the
coldest temperature in the minidisks, and therefore fall somewhere near the center of the range
of temperatures at which there is little thermal radiation:
\begin{equation}\label{eq:tempscale}
    T_0   = 3.3 \times 10^4 \left[\dot m (\eta/0.1)^{-1} M_8^{-1} (a/100R_g)^{-3}\right]^{1/4}\hbox{~K}\; .
\end{equation}
The accretion rate in Eddington units is $\dot m$, $\eta$ is the radiative efficiency, $M_8$
is the binary mass in units of $10^8 M_{\odot}$, and we scale the semi-major axis to a fiducial
value of $100R_g$ for reasons that will be explained in Section~\ref{sec:scalings}. Thus, for our
fiducial values, the spectral notch is cut in the near UV.

When the orbital evolution time of the binary (e.g., from gravitational
wave radiation) becomes shorter than the typical inflow time near the
inner edge of the circumbinary disk, most of the mass of the disk cannot
follow the binary compression inward to $r \simeq 2a$ \citep{MilosESP05};
we discuss this regime below.   First, however, we will discuss the case of larger
separations, which is likely to apply to many more objects.   So long as the
inner edge is located at $r \simeq 2a$, the temperature at this edge is
\begin{equation}
T_{\rm ie} = 2^{-3/4}T_0.
\end{equation}

In a binary with a circular orbit,  the tidal truncation radius data of \citet{Paczynski77} 
can be summarized by a simple fitting formula.   The truncation radius of the primary's
disk is at $R_1 \simeq 0.27 q^{-0.3} a$ and the truncation radius of the secondary's disk
is $R_2 \simeq 0.27 q^{0.3}a$.   We note further that these
fitting formulae imply that in units of the gravitational radii of the
individual black holes, $R_1/R_{g1} \simeq 0.27 q^{-0.3}(1+q)(a/R_g)$ and
$R_2/R_{g2} \simeq 0.27 q^{-0.7}(1+q)(a/R_g)$. Thus, these radii measured
in terms of the individual black holes increase by an amount depending on $q$.

Let the accretion rates onto the individual black holes be $\dot M_1
= f_1 \dot M$ and $\dot M_2 = f_2 \dot M$, with $f_1 + f_2 \lesssim
1$;  dynamical simulations indicate that, as might be expected from
the scale-free character of Newtonian gravity, $f_1$ and $f_2$ depend
only on the mass ratio $q$. According to \cite{Roedig12} and
\cite{Farris13}, in general $f_2 > f_1$. The temperatures at the
outside edges of the individual black hole disks are then
\begin{eqnarray}
T_1 = 0.27^{-3/4}\left(\frac{f_1}{1+q}\right)^{1/4} q^{9/40}T_0\\
T_2 =  0.27^{-3/4} \left(\frac{f_2}{1+q}\right)^{1/4} q^{1/40} T_0.\label{eq:temp3}
\end{eqnarray}
For $q=1$ and $f_1 = f_2 = 0.5$,
$T_1 = T_2 = 1.9 T_0$; in the limit as $q \rightarrow 0$ and $f_2 \rightarrow 1$, $T_2 \rightarrow 2.7 T_0q^{1/40}$.

Although we focus here on circular binaries, we note that \citet{2007ApJ...660.1624S} showed
that the tidal truncation radius for a binary with eccentricity $e$ is typically $\simeq (1-e)$ times the tidal
truncation radius for a circular binary of the same semimajor axis, to within $\sim 20$\%.
For analogous reasons, the inner edge of the circumbinary disk is likely to be pushed
outward.   Thus, in eccentric binaries we expect that the spectral notch will be wider than in
circular binaries, although eccentricity may also shift the lower edge of the notch to somewhat
higher energies because it leads to higher shock speeds when streams return to the inner edge.
The expected eccentricity depends on the separation and on which processes
dominate orbital evolution.  For gas-driven systems (in which gravitational radiation plays at most
a small role in driving coalescence), various studies suggest that the eccentricity could be as much as $\sim 0.6$ \citep{1991ApJ...370L..35A,2003ApJ...585.1024G,2005ApJ...634..921A,Cuadra2009,2011MNRAS.415.3033R,2012JPhCS.363a2035R}.  
On the other hand, when gravitational radiation dominates, binaries tend to circularize.  Indeed,
at our fiducial separation $\sim 100R_g$ (chosen for optimal observability, as we discuss later), binaries are likely
to be in the gravitational radiation-dominated regime.  We therefore expect
eccentricities to be relatively small and thus that the properties of
the notch will be fairly close to what they would be for circular binaries.

Provided the accretion rate is not a very small fraction of Eddington, either in the circumbinary
disk or in the minidisks individually, most annuli should be sufficiently dense and optically thick that the
local spectrum is well thermalized.
This is a generic property of all radiatively-efficient disks except in the very innermost rings of disks
accreting at near-Eddington rates, where both the density and optical depth fall to levels where thermalization
becomes marginal (\cite{K99}, Sec.~7.5.2).    In the present context, there may be deviations
at the various disk edges (the inner edge of the circumbinary disk, the outer edges of the mini-disks),
but we make the approximation that there is such a sharp surface density cut-off at these
edges that the disk is a thermal radiator on one side of the edge and dark on the other.
At the far outer edge of the circumbinary disk, low temperatures will likely make the opacity
absorption-dominated, but these regions radiate predominantly in the far-IR and produce
very little luminosity.
In the end, the great majority of the most relevant disk annuli are likely to be in a regime
in which electron scattering opacity strongly dominates absorption opacity at frequencies near
the peak of the local Planck spectrum.   When that is the case, the emergent spectrum may be
hardened by a factor $g \simeq 1.7$ \citep{ST95}.
Given the temperature estimated in Equation~(\ref{eq:tempscale}), this hardening
will apply to the circumbinary disk spectrum when $a \lesssim 1000 (\dot M/M_8)^{1/3} R_g$ and
to the individual disk spectra even for separations somewhat larger.   Thus, we expect that
the specific luminosity integrated from each of the three disks (the circumbinary disk and the two
minidisks) can be described by
\begin{equation}
L_\epsilon = \frac{32\pi^2}{3}\frac{a^2}{c^2h^3}(kT_0)^{8/3}g^{-4} \epsilon^{1/3}
   \int_{u_l}^{u_h} \, du \, \frac{u^{5/3}}{e^{u/g} - 1},
\end{equation}
where $u_l, u_h$ are lower and upper bounds to the integral corresponding to $h\nu/kT$ at the inner
($u_l$) and outer ($u_h$) radial boundaries appropriate to each disk.

At low energies, the luminosity per unit energy is dominated by the circumbinary disk because
the minidisk contributions are far down on the Rayleigh-Jeans tail of even the coolest portions of those disks.
It should therefore take the usual accretion disk form, but with an inner edge at $R_{\rm ie} \simeq 2a$ and an
outer edge effectively at $r = \infty$.   Thus, $u_l = u_{\rm ie} =  \epsilon/kT_{\rm ie} = 8^{1/4}\epsilon/kT_0$,
while $u_h = \infty$.    For $\epsilon \gtrsim g kT_{\rm ie}$, only the portion of the integral corresponding
to the Wien tail contributes; as a result, at these energies, the spectrum suffers an exponential cut-off .

At energies above this cut-off, the two inner disks dominate:
\begin{eqnarray} 
L_\epsilon^{(1)} &= (32\pi^2/3) (a^2 /c^2h^3) (kT_0)^{8/3} f_1^{2/3} (1+q)^{-2/3} g^{-4} \epsilon^{1/3}
   \int_0^{u_1} \, du \, \frac{u^{5/3}}{e^{u/g} - 1}\\
L_\epsilon^{(2)} &= (32\pi^2/3) (a^2 /c^2h^3) (kT_0)^{8/3} f_2^{2/3}  (1+q)^{-2/3} q^{2/3} g^{-4} \epsilon^{1/3}
   \int_0^{u_2} \, du \, \frac{u^{5/3}}{e^{u/g} - 1}.
\end{eqnarray}
By setting $u_l = 0$ in these two integrals, we have made the approximation that the
photon energies of interest are well below the temperature near the ISCO of the inner disks,
so that the bulk of the flux in this band is radiated by the outer regions of the inner disks.

The upper limits on the integrals are $u_{1,2} = \epsilon/ kT(R_{1,2})$, or
\begin{eqnarray}
u_1 &= 0.27^{3/4} {\epsilon \over kT_0}  f_1^{-1/4} (1+q)^{1/4} q^{-9/40}\\
u_2 &= 0.27^{3/4} {\epsilon \over kT_0}  f_2^{-1/4} (1+q)^{1/4} q^{-1/40}.
\end{eqnarray}
Immediately above the cut-off, the spectrum revives approximately $\propto \epsilon^2$
because the Rayleigh-Jeans tails of the thermal spectra from the two inner disks dominate.
Once energies comparable to $T_1$ and $T_2$ are reached, the classical $\epsilon^{1/3}$
spectrum is recovered.

\begin{figure}
\begin{center}
\includegraphics[width=0.6\textwidth,angle=90]{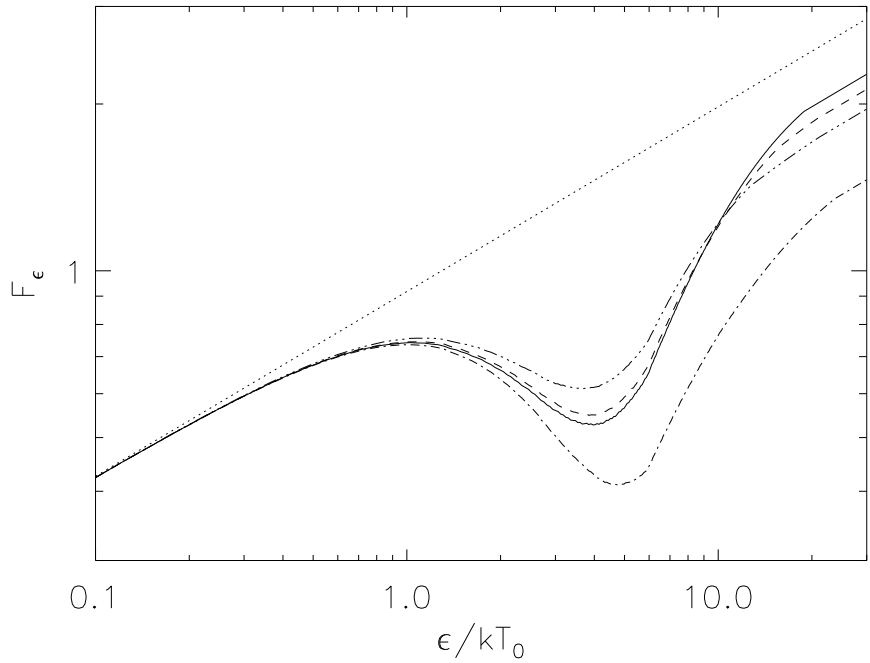} \\
\caption{Thermal spectra for four examples.   For all cases, the photon energy is in units of
$kT_0$; the hardening parameter $g=1.7$; and all curves share the same arbitrary flux units.
The solid curve shows the equal-mass case, $f_1 = f_2 = 0.5$ and $q=1$; the dashed curve shows
the case of $q=0.3$ and $f_1 = 0.45$, $f_2 = 0.55$; the two dash-dot curves show the spectrum
when $q=0.1$, and $f_1 = 0.08$, $f_2 = 0.92$ (lower curve) or $f_1=0.45$, $f_2=0.55$ (upper curve).}
\label{fig:farris}
\end{center}
\end{figure}

These analytic descriptions are illustrated in Figure~\ref{fig:farris}.   All three curves in
the figure assume: the circumbinary disk is in inflow equilibrium, so that $f_1 + f_2 = 1$;
the hardening factor $g=1.7$; and the maximum temperature for both inner disks is so high that
the spectrum does not reach its peak within the energy range shown (note that measuring
the radius of the secondary's disk edge in terms of its own gravitational radius, $R_2/R_{g2} \simeq 0.27 q^{-0.7}(1+q)(a/R_g)$,
so that $R_2/R_{g2}$ can actually be {\it greater} than $a/R_g$ when $q \ll 1$; thus, $R_2$
can be well outside the ISCO even for relatively small values of $a/R_g$).  
The solid curve illustrates the predicted spectrum for an equal-mass, circular orbit binary.    By symmetry,
$f_1 = f_2 = 0.5$.   The dashed curve shows the predicted spectrum for $q=0.3$ and the lower dot-dashed
curve for $q=0.1$ if $f_1/f_2$ is the value suggested by \cite{Farris13} but their sum has been adjusted so that
$f_1 + f_2 = 1$.    However, these relative accretion fractions remain rather tentative, so it is also useful to
explore the sensitivity of the spectra to their values.  The upper dot-dashed curve shows how things would
look if for $q=0.1$ the accretion fractions were the same as for $q=0.3$.

As can be seen from the figure, in this range of photon energies all the spectra lie well below the dotted line
illustrating the $\epsilon^{1/3}$ spectrum of a disk around a single black hole with the total
mass of the system.   The deep depression in the binary spectrum occurs because of the radiation that isn't emitted
between $\sim a/2$ and $2a$.   However, a weaker depression persists to higher energies.
The origin of this weaker depression is seen most clearly in the limit of $q \ll 1$, at which $f_2$ (as derived
from the Farris et~al. data) is $\simeq 1$.
In this situation, the temperature near the secondary's ISCO is higher than the temperature would be near the ISCO of a single
black hole because its mass is smaller, and consequently its spectrum extends to higher energies.
At the same time, however, the secondary's luminosity is almost as great as the total luminosity.
Combining these two effects, the binary spectrum must lie below that of the single-mass
case at energies less than the single-mass cut-off.

Comparing the binary spectra, we see that the spectrum changes very little for $0.3 \lesssim q \lesssim 1$.
Over this range in mass ratios, the specific flux in the deepest portion of the notch is a factor $\simeq 2.5$
below the spectrum of the single mass case, and this deepest depression appears at $\epsilon
\simeq 4kT_0$.   Recovery toward the single black hole level is slow because the inner disks have
smaller radiating areas than a disk around a single black hole at the same temperature.   Consequently,
the spectral depression lasts from $\simeq kT_0$ up to $\simeq 15kT_0$.

When the mass ratio falls below $q \simeq 0.3$, the spectrum may change more sharply if $f_2$
becomes as large as suggested by \cite{Farris13}.   It is then the case that the more massive black hole receives such a
small part of the accretion flow that its disk hardly contributes.    On the other hand,
because the tidal truncation radius for the secondary's disk has a radius $\propto q^{0.3}$, its radiating area
continues to decline as $q$ diminishes, even while the temperature at the truncation radius
saturates at $\simeq 2.7T_0$.    For these more extreme mass ratios and accretion fractions, the deepest part of
the notch is lower than for more moderate mass ratios and occurs at slightly higher energy.
In addition, the high energy continuum is depressed below the single black hole level by a factor
$\simeq 2$ until the cut-off temperature for the single black hole is reached.   On the other hand, if
for $q \simeq 0.1$, the ratio $f_2/f_1$ remains only somewhat greater than unity, the notch spectrum more
closely resembles the predictions for $q=1$ and $q=0.3$ because the primary continues to account
for a significant flux.

The breadth of the depression has a significant observational implication: it will be hard to
see its entire span in any single spectrum.   Ground-based observations have at most
a spectral range of a factor of 2; from the atmospheric cut-off to the Lyman edge is only
a factor of 4; these notches can be expected to span a factor of 10 or more.

\begin{figure}
\begin{center}
\includegraphics[width=0.6\textwidth,angle=90]{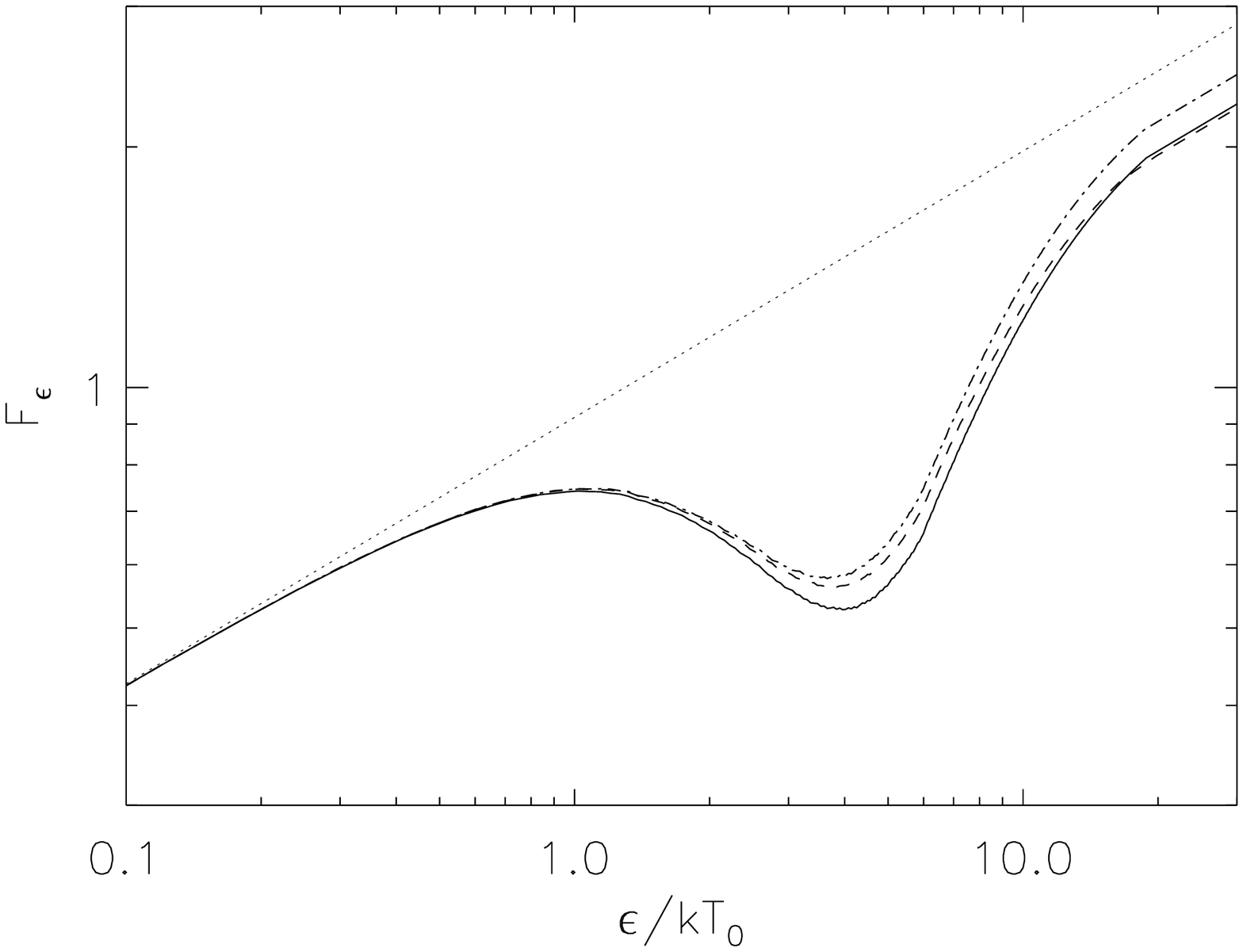} \\
\caption{Thermal spectra like those of Fig~\ref{fig:farris} but with accretion fractions
$f_1,f_2$ exchanged.  The units and other assumptions are the same as those in Fig.~\ref{fig:farris}.
The solid curve shows the equal-mass case, $f_1 = f_2 = 0.5$ and $q=1$; the dashed curve shows
the case of $q=0.3$ and $f_1 = 0.55$, $f_2 = 0.45$;
the dash-dot curve shows the spectrum when $q=0.1$ and $f_1 = 0.92$, $f_2 = 0.08$.}
\label{fig:antifarris}
\end{center}
\end{figure}

As we have seen already, the specific predictions for how the shape of the notch depends on $q$ are closely tied
to the dependence of the accretion fractions on that quantity.  To underscore further the importance
of firming up this relationship, in Figure~\ref{fig:antifarris} we show what the spectra would be if the
mass fractions based on the Farris et~al. results were reversed.   If this were how the accretion rates were divided,
the notch would weaken as $q$ becomes smaller; to the extent the primary acquires most
of the total accretion rate, its spectrum more and more resembles the single-mass limit.

All of these remarks pertain to the regime in which the ratio $R_{\rm ie}/a$ depends
only on $q$ (and rather weakly).
Late in the evolution of a black hole binary, gravitational
wave radiation accelerates orbital shrinkage, and eventually the orbital evolution time becomes
shorter than the inflow time near the inner edge of the circumbinary disk \citep{MilosESP05}.   From that point
onward, $R_{\rm ie}$ remains fixed, although the shape of the edge is likely to become less
abrupt as a low surface density ``foot" stretches into the gap \citep{Noble12}.   On the other
hand, $R_{1,2}$ continue to shrink in proportion to $a$.   Even though the disk cannot
transport the bulk of its material inward fast enough to keep up with the diminishing binary
separation, it nonetheless can continue to leak matter into the gap at a rate similar to its
intrinsic accretion rate \citep{Noble12}; this is possible because the total mass accreted during
this period of binary evolution is, by definition, only a small fraction of the mass near the
circumbinary disk's inner edge.

Thus, the situation during this stage is that $T_{\rm ie}$ changes very little, but $T_{1,2}$ rise
while $R_{1,2}$ fall.    As a result, the low-energy onset of the notch does not move, but the
energy at which the thermal spectrum recovers moves higher, broadening the notch's width.
Ultimately, when $R_{1,2}$ become comparable to their respective ISCO radii, the minidisks
disappear and the thermal continuum never does recover at high energies.    From this point
to black hole merger is only a short time.

\section{Hot-spot radiation}

Matter moves from the inner edge of the circumbinary disk to the outer edges of the
individual black hole disks on ballistic orbits with little dissipation {\it en route}
\citep{MM08,Roedig12,Shi2012,Noble12,Farris13}.    As the gas accelerates inward,
adiabatic expansion keeps the temperature of the streams less than what it was upon
departing from the circumbinary disk.   Because both the stream speed and the orbital
motions in the circumbinary disk are in most cases
highly supersonic, the streams shock when they hit the individual disk edges.
The relative height of the disk proper vis-a-vis the vertical thickness of the stream
does not matter for the shock location because, as we will show shortly, these
shocks create an extremely hot annulus around the disk edge, and the shock front
occurs at the outer edge of these hot, and therefore vertically thick, annuli.

The relative velocity between an incoming stream and the matter orbiting at the outer
edge of an individual disk is ${\bf v}_{\rm stream} - {\bf v}_{1,2} - {\bf v}_{\rm orb,1,2}$, where
${\bf v}_{\rm stream}$ is the stream velocity with respect to the binary center of mass, ${\bf v}_{1,2}$
is the velocity in the binary center-of-mass frame of either the primary or the secondary,
and ${\bf v}_{\rm orb,1,2}$ is the
orbital velocity of material around either black hole at its disk's outer edge.  The shock
speed is the component of this relative velocity in the direction to the individual black hole.

One might guess that the magnitude of the shock speed would be comparable to ${\bf v}_{\rm orb}$.
To test this guess and refine the estimate, we have used data from the circumbinary disk
simulations of \cite{Roedig12} to define the initial conditions of test-particles traveling from
the inner edge of the circumbinary disk to the edge of the tidal truncation radii around a
pair of black holes with mass ratio $q = 1/3$ traversing orbits of eccentricity $e = 0$,
0.2, 0.4, 0.6, and 0.8.   We then found the mean shock speed for each black hole disk
at each eccentricity by averaging over these distribution functions.   Although fluid
effects can alter these velocities, direct evaluation of hydrodynamic (and magnetohydrodynamic)
forces relative to gravity (e.g., \citealt{Shi2012}) indicate that the fluid effects enter only
at the $\sim 10\%$ level.  For this mass ratio, we find that the shock speed at the primary's
disk increases from $\simeq 0.9 v_{\rm orb,1}$ to $\simeq 2.1 v_{\rm orb,1}$ as
the eccentricity increases from 0 to 0.8, with most of the change occurring at larger
eccentricity.  This comparative insensitivity to eccentricity is the result of the primary
staying close to the system center of mass at all times and all eccentricities.

At the secondary's disk, on the other hand, the shock speed 
 is rather more strongly dependent on eccentricity.  It increases
from $\simeq 0.6 v_{\rm orb,2}$ for $e=0$ to $\simeq 3 v_{\rm orb,2}$ at $e=0.8$.
Guided by these results, we will parameterize our results here in terms of an order-unity
multiple $\Phi$ of the appropriate orbital speed.   For a circular orbit binary, $\Phi_1 \simeq 0.9$,
while $\Phi_2 \simeq 0.6$.

In those terms, the shock speeds are
\begin{equation}
v_{s1,2} = 0.19 c \Phi_{1,2} (q^{0.15},q^{0.35})(1+q)^{-1/2} (a/100R_g)^{-1/2},
\end{equation}
where the notation $(x,y)$ means that $x$ applies for the primary and $y$ for the secondary. 
If the post-shock gas is dominated by relativistic particles, whether photons or electrons,
its adiabatic index $\gamma = 4/3$.   The immediate post-shock temperatures are then
\begin{equation}
T_{s1,2} = 6.2 \times 10^{10} [(1.4 + 1.2Zm_e/m_p)/(1.1 + 1.2Z)] (a/100R_g)^{-1} \Phi^2_{1,2}(1+q)^{-1} (q^{0.3},q^{0.7}) \hbox{~K}.
\end{equation}
Here $Z$ gives the ratio of total electrons (i.e., including positrons) to net leptons.
This temperature is in fact likely to be an upper bound.  At temperatures $\gtrsim 10^9$~K, there
can be rapid pair production; for fixed internal energy, pair production leads to a smaller effective
adiabatic index, decreasing the temperature both through the ``latent heat" of pair production (the
rest-mass energy) and through dividing the internal energy over a larger number of particles.   In addition,
as we will discuss in more detail, radiative cooling can be quite rapid.   Nonetheless,
because the nominal post-shock temperature is comparable to the virial temperature and there
must be a very large amount of pair production before the rest-mass density is significantly increased,
the post-shock gas should have a scale height $h_{1,2} \lesssim R_{1,2}$.

The gas density in the  streams immediately upstream of the shock is
\begin{eqnarray}
\rho_{1,2} = 1.4 \times 10^{-14} &\dot m  M_8^{-1} (\Delta\phi/2\pi)^{-1} (\eta/0.1)^{-1} (a/100R_g)^{-3/2} (1+q)^{1/2}\nonumber \\
          &\quad \times f_{1,2}(h_{1,2}/R_{1,2})^{-1} \Phi^{-1}_{1,2} (q^{0.45},q^{-0.95}) \hbox{~g~cm$^{-3}$}.
\end{eqnarray}
Here $h_{1,2}$ are the vertical thicknesses of the streams at this point and $\Delta\phi$, which may
be $\ll 2\pi$, is the azimuthal extent of the stream impact.    Note that the radiative efficiency of the minidisk
appears in the gas density rather than the photon energy density because we are defining
$\dot m = L/L_E$.   As a
result of the shock, the density is increased by a factor $D$, which is 7 if
$\gamma = 4/3$.   If the radial thickness of the hot gas annulus is $\Delta r$,
the mass in the annulus must be exactly the mass brought to the disk during
a hot annulus cooling time $t_{\rm cool}$.    Consequently,
\begin{equation}
 \Delta r = v_{s} t_{\rm cool}/D,
\end{equation} 
where we suppose, as we will show shortly, that $t_{\rm cool} \ll \Omega^{-1}$ so that
there is too little time for the shocked gas to spread either azimuthally or vertically.
When this assumption is correct, $\Delta r \ll R_{1,2}$ because $v_s \sim v_{\rm orb}$
and $D > 1$.

At such high temperatures, the two radiation mechanisms most likely to be relevant
are bremsstrahlung (possibly in the relativistic regime) and inverse Compton scattering
on the thermal photons produced by the individual minidisk whose hot annulus is under
consideration.   In many other circumstances, the bremsstrahlung luminosity is
proportional to one more power of the density than the inverse Compton luminosity;
here, however, because the seed photon intensity is {\it also} proportional to the
gas density through their common tie to the accretion rate, the ratio of the two is
independent of the density (and therefore the accretion rate).

In this context it is therefore convenient to write the bremsstrahlung emissivity $j_{\rm ff}$ in the
form $n_i n_e\alpha_{fs}\sigma_T m_e c^3 f_b(\Theta)$, where $\alpha_{fs}$ is the
fine-structure constant and $\Theta$ is the electron
temperature in rest-mass units; in the non-relativistic limit, $f_b(\Theta) = g(\Theta)\Theta^{1/2}$,
where $g(\Theta)$ is a slowly-varying Gaunt factor (Krolik 1999).   We will
similarly write the Compton emissivity $j_{\rm C}$ as $n_e \sigma_T c U_\gamma f_C(\Theta)$,
where $U_\gamma$ is the photon energy density and $f_C(\Theta) = 4\Theta$ in the
non-relativistic limit.   Their ratio is then
\begin{equation}
\frac{j_{\rm ff}}{j_{\rm C}} = \frac{n_i \alpha_{fs} m_e c^2 f_b(\Theta)}{U_\gamma f_C(\Theta)}.
\end{equation}
In other words, in the regime for which the functions $f_b$ and $f_C$ are both
order unity, the bremsstrahlung emissivity is greater than the Compton emissivity if
the rest-mass density of the net number of electrons exceeds $U_\gamma/\alpha_{fs}$.

Both the ion density and the photon density are proportional to the accretion rate onto
a given black hole because the seed photons are themselves created by that accretion.
However, the rate at which seed photons are produced by the inner disk is also
dependent on an efficiency, a factor which does not enter the density.
As a result, the bremsstrahlung-Compton ratio is inversely proportional to the disk's
radiative efficiency but independent of accretion rate.   It is also independent of
the number of electron-positron pairs because both the free-free and the Compton
emissivity are proportional to the total number of electrons:
\begin{eqnarray}
\frac{j_{\rm ff}}{j_{\rm C}} = 52 & (2\pi D/\Delta\phi) \alpha_{fs} (m_e/\mu_i)[f_b(\Theta)/f_C(\Theta)] (\eta/0.1)^{-1} (a/100R_g)^{1/2} (h_{1,2}/R_{1,2})^{-1}\nonumber\\
            &\quad \times \Phi_{1,2}^{-1}(1+q)^{1/2}(q^{-0.15},q^{-0.35}).
\end{eqnarray}
Thus, to order of magnitude, this ratio is a product of atomic constants ($\mu_i$ is
the mass per ion) that is always $\sim 10^{-5}$, a function of the mass ratio generally not far
from unity, and a parameter ratio ( $52 (2\pi D/\Delta\phi) (R_{1,2}/h_{1,2})/\Phi_{1,2}$ )
that may be large, but in most instances not large enough to make
the overall product comparable to unity.  The fact that we have neglected Compton
amplification of $U_\gamma$ also means that this is a conservative conclusion;
any amplification further strengthens Compton losses relative to free-free.

If inverse Compton scattering does dominate the cooling rate, the cooling time will be (in units of
the dynamical time)
\begin{equation}
t_{\rm cool}\Omega \simeq 3.5 \times 10^{-2} (1 + 1.1/Z)  (\dot m f_{1,2})^{-1} \Phi_{1,2}^2 
\frac{\Theta}{f_C(\Theta)}(a/100R_g)^{-1/2}(1+q)^{-3/2} (q^{0.15},q^{1.35}).
\end{equation}
Thus, the cooling time will typically be shorter than an orbital period, although it might become
comparable to the orbital period when $a \ll 100R_g$ or $\dot m f_{1,2}\ll 1$.   The shock on the
secondary's disk cools especially rapidly when $q \ll 1$.   The net result is that
the hot shocked region should form an incomplete annulus around the outer edge of the
individual disk, and essentially all the potential energy made available by the streams' ballistic
fall from the inner edge of the circumbinary disk to the outer edge of an individual disk is
radiated in these annuli.

Because essentially all the heat created by the shock is radiated quickly in the hot spot, the luminosity is
\begin{equation}
L_{\rm hot} = (3/2)\left(1 + 1.1Z\right)\left(kT_{s1} f_1 + kT_{s2}f_2\right)\dot m (L_E (M)/c^2\eta\mu_i).
\end{equation}
When Compton scattering dominates, as it generally should, the shape of the spectrum can be described
phenomenologically by the Compton-$y$ parameter ($\equiv 4\Theta \tau_T(1 + \tau_T)$ for Compton
optical depth $\tau_T$)  and the temperature if the cooling timescale is long compared to the photon escape
timescale .   We have already estimated an upper bound for the post-shock temperature; the Compton optical depth is
\begin{equation}
\tau_T \simeq  6.7 Z (\dot m f_{1,2}/\Phi_{1,2})(D/7)(a/100R_g)^{-1/2}(\eta/0.1)^{-1}(1+q)^{1/2} (q^{0.15},q^{-0.65}) .
\end{equation}
As this expression shows, for fiducial values of the parameters, $\tau_T \gtrsim 1$ is to be expected,
particularly if there is copious pair production, but smaller accretion rates could diminish the optical depth. 
The Compton-$y$ parameter is then
\begin{equation}
y = 45 \Theta Z^2 (\dot m f_{1,2}/\Phi_{1,2})^2 (D/7)^2 (a/100R_g)^{-1}(\eta/0.1)^{-2}(1+q) (q^{0.3},q^{-1.3}) .
\end{equation}
If $\Theta$ is given by the upper bound for the post-shock temperature,
\begin{equation}
y \simeq 180 Z^2 (\dot m f_{1,2}\Phi_{1,2})^2 (D/7)^2 
                          (a/100R_g)^{-1}(\eta/0.1)^{-2}(1+q) (q^{0.3},q^{-1.3}) .
\end{equation}
Thus, for accretion rates not too much smaller than Eddington, Comptonization
will be very strong (again particularly if there is a great deal of pair production)
and a Wien-like spectrum will emerge with a characteristic temperature set
by the electron temperature $T_e$.
Comptonization will be especially strong in the secondary's hot spot if $q \ll 1$.
Note that if the shock were at the much smaller radius suggested by \cite{Tanaka13b}, the gas density
and optical depth would be so much higher that the emergent spectrum would be thermal, more nearly 
resembling the spectrum we predict emerges from the inner portions of the mini-disks.

As we remarked earlier, $T_{s1,2}$ is the highest temperature the radiating electrons may reach.  A
number of effects may lower it.    It is possible, for example, that if the dissipation in the shock primarily
goes into the ions rather than the electrons, heat transfer from the ions to the electrons may be slow
enough that Compton cooling will keep the electron temperature below the ion temperature.   If so,
this effect would also extend the effective cooling time of the shocked gas and widen the hot annulus.
One measure of the rate of ion-electron heat transfer is the time $t_{\rm ion}$ required to change the
ion temperature by Coulomb collisions \citep{GS85,Mahadevan97}.  Normalizing this rate per ion in the
same units in which we previously evaluated the cooling time, it is
\begin{eqnarray}
t_{\rm ion}\Omega = &0.64 (1 - T_i/T_e)^{-1}  (D/7)^{-1} [Zg(\Theta)]^{-1} (\Delta \phi/2\pi) (h/R)_{1,2} \nonumber\\
      &\quad \times (\dot m f_{1,2})^{-1} (\eta/0.1)  (a/100R_g)^{1/2} \Phi_{1,2} (1+q)^{-1/2}(1,q^{1/2}).
\end{eqnarray}
Thus, even if most of the heat generated in the shock is given to the ions, and Coulomb collisions are the only
heat exchange mechanism, because $\Delta\phi \ll 2\pi$ and $h/R \ll 1$, ion cooling will in most circumstances
be rapid compared to the orbital frequency.  It is possible, however, that it may not be rapid with respect to the
Compton cooling rate, especially if $\dot m f_{1,2} \ll 1$.   On the other hand, given the level of plasma turbulence
one might expect in a high Mach-number shock, other ion-electron heat transfer mechanisms might also
operate \citep{Sharma07} that would allow ion-electron heat exchange to keep up with Compton cooling.

Strong pair production would reinforce this conclusion.     The importance
of this process is conveniently parameterized by the compactness $\ell
\equiv L_{\rm hot}\sigma_T/(r m_e c^3)$.   Here $r$ is the characteristic
length scale of the plasma, which we take to be $\sim
R_{1,2}\Delta\phi_{1,2}$.    With this guess for the length scale, we
find \begin{equation} \ell_{\rm hot} \simeq 6 \dot m f_{1,2}
(2\pi/\Delta\phi) \Phi_{1,2}^2 (a/100 R_g)^{-2} (1+q) (q^{0.6},q^{0.4}).
\end{equation} Provided $\Delta\phi \ll 2\pi$, the compactness could be
large.  Strong pair production can be expected when $\ell \gtrsim 1$ and
the electron temperature is driven to $\gtrsim 10^9$~K.    As noted
earlier, when pair production is strong, it also tends to cap the electron
temperature in this range \citep{Pietrini95}.

However, all these estimates must be taken as rather tentative because the post-shock plasma
is so far from equilibrium.  The heating is almost
instantaneous relative to the cooling time, and the pair creation time is likely
comparable to the cooling time.   A detailed time-dependent pair-balance,
thermal-balance, and Compton-scattering calculation would therefore be necessary
for a proper prediction of the spectrum.

Further complications arise in the decoupling regime.   When the
minidisks shrink with the shrinking binary orbit, but $R_{\rm ie}$ does
not change, the shock temperatures rise as $R_{1,2}$ become smaller, and
the luminosity rises in proportion.   The compactness also grows, so
that pairs can be expected to be still more important.  When the specific
angular momenta of the accretion streams become smaller than the specific
angular momentum at the ISCO, the disks disappear and so do the hot spots.

\section{Observability}\label{sec:scalings}

The observability of the notch and the Wien-like spectrum depend on 
several considerations, including the band in which the notch occurs,
the luminosity at wavelengths near the notch, the luminosity of the hot spot,
and the longevity of the particular situation.   Each of these in turn depends
on parameters such as the accretion rate, the total mass of the binary, the
binary mass ratio, and the binary separation.

As illustrated by Figure~\ref{fig:farris}, we estimate that the minimum of the 
notch region falls at $\epsilon_n \simeq 4kT_0$, with its low-energy edge
at $\simeq 1.5 kT_0$ and its high-energy recovery at $\simeq 15kT_0$.
If we were to require that the minimum of the notch fall in the middle of the visible band,
that would impose the constraint
\begin{equation}
\dot m (\eta/0.1)^{-1} M_8^{-1} (a/100R_g)^{-3} \simeq 2 \times 10^{-3}(1+z)^4.
\end{equation}
To place the low-energy end of the notch at those wavelengths would make
the constraint on the parameter product $\simeq 50\times$ smaller;
conversely, to place the high-energy end of the notch at this wavelength
would enlarge the constraint on the parameter product by a factor $\simeq
40$. Given that telescopic sensitivity is maximal in the optical/NIR
band, this means that observability of the notch (in wavelength terms) is maximized when the
separation is  several times greater than our fiducial value of
$100~R_g$, or the accretion rate is well below Eddington, or the total
mass is relatively large.   The notch is also more likely to fall in the optical band
when the source is at high redshift.   Instruments sensitive in the UV, so that the constraint
is based on a shorter wavelength, would, of course, make it easier to satisfy.
However, as we discuss below, when the notch is at short wavelengths or
high redshifts, the lifetime of the binary against gravitational wave-driven coalescence is short.

Even when the wavelength of the notch occurs at an easily-observable
wavelength, detectability is enhanced, of course, by greater flux.   If the disk
were uninterrupted, the luminosity in the notch region would be $L_n
\sim \dot m L_E (r_m/a)$, where $r_m$ is the radius of maximal surface brightness,
generally $\simeq 1.3\times$ the radius of the ISCO \citep{Noble11}.
Because neighboring spectral regions have comparable luminosity, one
way of phrasing the detectability criterion is that $L_n \gtrsim L_*$, where $L_*$
is the usual characteristic luminosity of the galaxy luminosity function in the
rest-frame band of the notch, $\simeq {\rm several} \times 10^{43}~{\rm erg~s}^{-1}$, depending on
which band that is (see \citealt{2011A&ARv..19...41J} for a recent review).   That
criterion is
\begin{equation}
\dot m M_8 (a/100R_g)^{-1} (r_m/10R_g) \gtrsim 0.05,
\end{equation}
Thus, sources with masses comparable to or even somewhat smaller than
our fiducial value, or accretion rates not too small compared to Eddington, should be
at least  as bright in the ``notch band" as $L_*$ galaxies.   They should
therefore be easily detectable in photometry out to quite high redshift.

The luminosity scale also affects the observability of the hard X-rays produced by the
hot spot.   At the order of magnitude level, $L_{\rm hot} \simeq L_n$ because the energy available
to be radiated in the hot spot is simply the energy that is {\it not} radiated in the gap between
the inner edge of the circumbinary disk and the outer edges of the inner disks.   In this case,
however, the standard of comparison is the intrinsic hard X-ray spectrum radiated by the
disks' coron\ae\ near the black hole.   In ordinary AGN, the luminosity
at $\sim 100$~keV is typically a fraction $F \sim 0.2$ of the bolometric luminosity
\citep{2012ApJ...745..107W}.    Thus, the hot-spot X-rays are able to stand out clearly
against the ordinary coronal X-ray spectrum when $r_{\rm m}/a \gtrsim F$.   More precisely,
if we approximate the hot-spots' luminosity by the binding energy liberated when the streams join
circular orbits of radius $R_{1,2}$ around the black holes, the ratio of total hot-spot luminosity
to coronal luminosity is approximately
\begin{equation}\label{eq:hotspotratio}
L_{\rm hot}/L_X \simeq 0.35 (F/0.2)^{-1} (a/100R_g)^{-1} (1+q)^{-1} (\eta/0.1)^{-1}
     \left[ f_1 q^{0.3} + f_2 q^{0.7}\right].
\end{equation}
The magnitude of this ratio is one of the motivations for our choice of
$100R_g$ as the fiducial scale for the separation.

\begin{figure}
\begin{center}
\includegraphics[width=0.6\textwidth,angle=90]{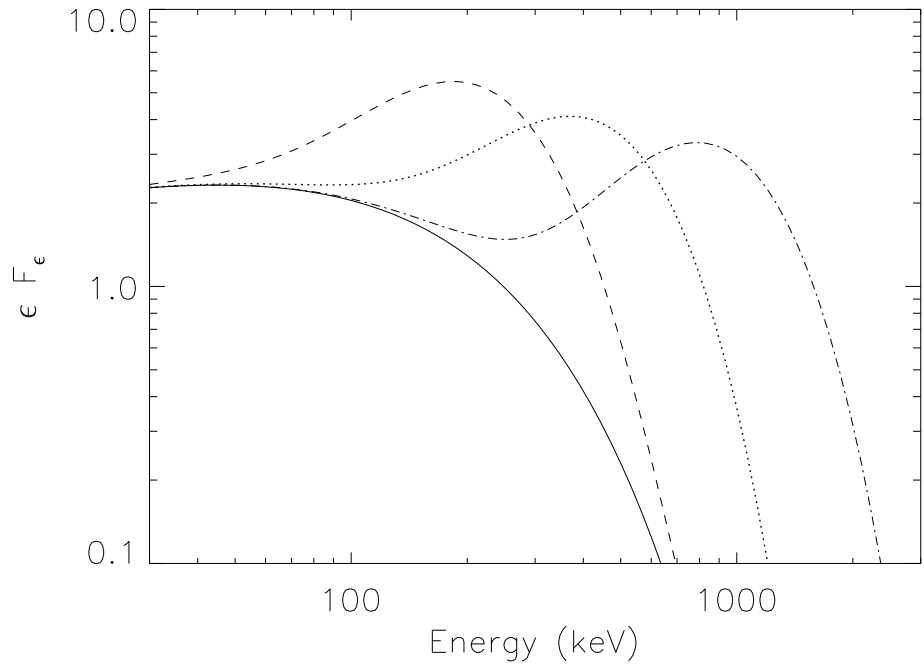} \\
\caption{Four possible hard X-ray spectra: that of the type~1 Seyfert galaxy
IC4329A, using the 6--70~keV fit to {\it NuStar} data \citep{Brenneman13} (solid curve); and three Wien spectra with
temperatures 100~keV (dotted curve), 50~keV (dashed curve), and 200~keV (dot-dash
curve) and luminosity equal to that in the IC4329A spectrum added to the IC4329A spectrum.}
\label{fig:hotspot}
\end{center}
\end{figure}

However, just as for the notch, the non-uniform sensitivity of available
instrumentation makes the ease of detecting the hot-spot radiation depend
upon the specific parameters governing its characteristic temperature.
Although the general characteristics of Compton-cooling and pair-producing
plasmas lead to temperatures $\gtrsim 10^9$~K, factors of a few matter
here, and an effort of larger scope will be required to refine this estimated
temperature.   Consequently, here we will confine ourselves to some
qualitative remarks.   First, because $\epsilon F_\epsilon \propto \epsilon^4 e^{-\epsilon/kT_e}$
for a Wien spectrum, the peak in such a spectrum falls at $\simeq 4 kT_e$.
Second, to illustrate the potential impact of the actual characteristic temperature
of the radiated spectrum, we present in Figure~\ref{fig:hotspot} three sample spectra
constructed by adding hot-spot spectra with different temperatures
to a ``typical" coronal X-ray spectrum from a conventional
AGN (in this case, IC4329A: \citealt{Brenneman13}).    In each case, the luminosity
in the hot spot radiation is chosen to be equal to the hard X-ray luminosity of the
conventional AGN spectrum.   As these examples show, a true Wien spectrum
is so hard that when added to a conventional spectrum it produces a very distinct
peak.    However, as these examples also show, the Wien peak is likely to be found
above 100~keV, so that, given the limitations of hard X-ray instrumentation,
detections are more likely to rely on a spectral hardening at energies $\lesssim 100$~keV.

This estimate of the hot-spot's observability depends upon measuring the time-averaged
hard X-ray spectrum of the source.   It is possible that the hot-spot emission is modulated strongly
at frequencies comparable to the binary orbital frequency \citep{Shi2012}.   If so, it might be possible
to detect the hot spot even when $L_{\rm hot}/L_X$ is rather smaller than unity by searching for
a periodic component in the flux.   This effort would require a sizable amount of observing time,
however, both for individual measurements and to construct a suitable time series.   In this
last regard, it is convenient that the orbital period in the most interesting regime is $\sim 0.1 M_8 (a/100R_g)^{3/2}$~yr.

The final consideration governing observability is the size of the population with
separations in the interesting range.    \cite{Sesana11} estimate that the number of
mergers in the universe per year with total binary mass $\sim 10^8 M_{\odot}$ is $\sim 1$, with the greatest
number of them at $z \simeq 3$--6 and having $q \sim 0.1$.   That figure can
be translated into a total population at larger separation by multiplying by the lifetime
at the separation of interest.   In the range of $a$ of interest to this paper, the evolution of the
binary is almost certainly dominated by energy loss in gravitational radiation, because the gravitational wave inspiral time is much less than the characteristic time for gas or stellar processes to affect the orbit.  More specifically, 
when the orbit is circular, the lifetime is
\begin{equation}
T_{\rm GW}={5\over{256}}{a^4c^5\over{\nu G^3M^3}}
\end{equation}
\citep{Peters1964}, where $\nu=M_1M_2/M^2$ is the symmetric mass ratio;
$\nu$ reaches its maximum of 1/4 when $M_1=M_2$. 
Combining the gravitational wave lifetime with the constraint that a given
part of the notch is in the middle of the visible band ($\lambda_* = 5000$~\AA), we find a lifetime for the
most readily-observable notch systems of
\begin{equation}\label{eq:GW}
T_{\rm GW}=2.2\times 10^3 {\dot m}^{4/3}(C/1.5)^{16/3}(\eta/0.1)^{-4/3} (\lambda_*/5000\hbox{\AA})^{16/3}
(1+z)^{-13/3} (4\nu)^{-1} M_8^{-1/3}\hbox{~yr}.
\end{equation}
Here $T_{\rm GW}$ is the lifetime in our frame, and $C\equiv
hc/(\lambda_*kT_0)$ is a measure of which part of the notch we are
considering.  From our previous discussion, if we assume a spectral
hardening factor $g=1.7$, then $C\approx 1.5$ at the low-energy edge of
the notch, $C\approx 4$ at the minimum, and $C\approx 15$ at the
high-energy edge of the notch.  Clearly, the expected lifetime is
extremely sensitive to the wavelength of observation, to the redshift,
and to the relevant portion of the notch.   For example, if 
$\lambda_* = 5000$~\AA, $z \simeq 4$, $q=0.1$, and $C=1.5$ then $T_{\rm
GW}\approx \sim 6 {\dot m}^{4/3}$~yr and hence we expect $\sim 6{\dot
m}^{4/3}$ such systems to be currently active.  If instead
$\lambda_*=8000$~\AA, $z \simeq 1$, $q=0.1$, and $C=4$, then $T_{\rm
GW}\approx 2.5\times 10^5 {\dot m}^{4/3}$~yr and hence there would be
potentially hundreds of thousands of such systems.   The fact that one plausible
set of parameters yields a population five orders of magnitude greater than another,
equally plausible, set of parameters underlines the large uncertainty in our 
estimate of the size of the observable population.

Applying a similar criterion to the hot-spot radiation, we might require $L_{\rm hot}/L_X \geq 0.5$
(cf. Equation~(\ref{eq:hotspotratio})).    If this were the only prerequisite for detection, the associated
lifetime would be
\begin{equation}
T_{\rm GW} \leq 30 (1+z)(F/0.2)^{-4}(1+q)^{-4}(\eta/0.1)^{-4}\left(f_1q^{0.3} + f_2q^{0.7}\right)^4M_8(4\nu)^{-1}\hbox{~yr}.
\end{equation}
The factors dependent on the mass-ratio have limits of 1/4 for $q=1$ and $q^{-0.3}$ for $q \ll 1$,
so that the lifetime increases slowly for smaller $q$.    The hot spot hard X-rays are therefore most readily visible
from systems anywhere from roughly one to several decades before merger.    Discovery of an example could then
be taken as an early warning of a merger whose gravitational waves might be detectable; knowledge of the precise
direction to the event could aid in parameter estimation from an indirect detection of the gravitational waves using a
pulsar timing array \citep{Sesana10}.
Normalizing to the event rate as we did for the notch, our estimate suggests some dozens in the sky at any given time.

Unfortunately, our ability to detect this hard X-ray signal depends strongly on the energy at which the feature may
be seen.   In this regard, redshift plays an ambiguous role.   It may shift the energy of the peak in
luminosity down to bands where instrumentation is more sensitive; on the other hand, for any
given luminosity, redshift also exacts a cost in reduced flux.

\section{Conclusions}

Previous work pointed out that the interruption of a circumbinary disk surrounding a pair
of black holes leads to a cut-off in its thermal continuum spectrum at much lower energies
than the cut-off expected in a disk around a single black hole of the same mass.   We
have shown that the spectrum should, in general, revive at energies $\sim 10\times$ greater
because the minidisks surrounding the individual black holes become bright at those
energies.   In addition, we have, for the first time, discussed how the properties of such a
spectral ``notch" depend upon system mass, mass ratio, and separation.

We have also introduced a new continuum spectral signature characteristic of black
hole binaries that will 
be most prominent at rest-frame energies of tens to hundreds of keV.  This
radiation comes from the hot spots produced when accretion streams fall
nearly ballistically from the inner edge of the circumbinary disk and
hit the outer edges of the minidisks around the individual supermassive
black holes.  These hot spots should radiate Wien spectra with temperatures
$\sim 100$~keV, although given the current limitations of hard X-ray detectors, these
spectra might be easier to detect as spectral hardening at tens of keV or above, or
possibly by a periodic modulation in the X-ray flux at those energies.

There are, however, certain difficulties associated with searches for either one.
The number of sources in the sky with a notch in the thermal continuum spectrum
at wavelength $\lambda_*$ is extremely sensitive to $\lambda_*$ itself ($\propto \lambda_*^{16/3}$)
as well as the redshift of the source ($\propto (1+z)^{-13/3}$) and even the
portion of the notch falling at $\lambda_*$ (there are $\sim 200\times$ more
systems at any given time for which the minimum of the notch is at $\lambda_*$
than those for which the low-frequency edge of the notch is at $\lambda_*$).
If these searches wish to use the high sensitivity of optical band instruments,
they will be most sensitive to systems in which the parameter combination
$\dot m M_8^{-1} (a/100R_g)^{-3} (1+z)^{-4} \sim 10^{-3}$.
Similarly, for the Wien bump of the hot-spots to stand out clearly against the coronal
hard X-rays generated by the usual processes in the innermost portions of
the minidisks, it is necessary for $(a/100R_g) \lesssim 1$, although it is possible
that modulation of the hot-spot luminosity at a period comparable to the orbital
frequency may relax this criterion somewhat.   Thus, these
signatures are particularly powerful for identifying supermassive black hole
binaries with separations $\sim 10^{2\pm 1}R_g$ or less.

In any event, both spectral signatures---the notch at longer wavelengths and the Wien bump
in the hard X-ray band---are distinctive signatures of accretion onto black hole
binaries.   It is hard to imagine another system that could produce either, much
less both, of these features.    Thus, detection of either one would represent a
strong suggestion, and detection of both a likely
discovery of a supermassive binary black hole system on its way toward merger.

\section*{Acknowledgements}

CR wishes to thank Kip Kuntz for his patient explanations of instrumental features and how to read the X-ray literature,
Guangtun Ben Zhu for his help using databases, Tim Heckman for guidance, and Roberto Decarli for an invitation to
and warm hospitality at the MPIA Heidelberg, where the idea for this paper originated.
We thank Jeremy Schnittman for helpful conversations.
This work was partially supported by National Science Foundation grant 
AST-1028111 (to JHK) and NASA ATFP grant NNX12AG29G (to MCM).  This work was
also supported in part by a grant from the Simons 
Foundation (grant number 230349 to MCM).   MCM thanks the Department 
of Physics and Astronomy at Johns Hopkins University for their hospitality 
during his sabbatical.

\bibliography{Bib}
\end{document}